\documentclass[3p,number,times,sort&compress]{elsarticle}

\usepackage{lineno}
\usepackage[colorlinks,
            linkcolor=blue, 
            anchorcolor=blue,
            citecolor=blue, 
            ]{hyperref}

\usepackage{color}
\usepackage{amsmath}
\modulolinenumbers[1]
\usepackage[utf8]{inputenc}
\usepackage{graphicx}
\usepackage{amsmath}
\usepackage{booktabs}
\usepackage{subfigure}

\usepackage{amssymb}
\usepackage{ulem}
\usepackage{multirow}
\usepackage{soul}
\setstcolor{red}

\usepackage{arydshln}
\usepackage{microtype}

\journal{JINST}

\begin{document}

\begin{frontmatter}

\title{A Multimodal Domain-Adversarial Network for Fragmentation Background Suppression in AMS Heavy Nuclei Measurements}

\cortext[mail]{Corresponding author}

\author[firstaddress,secondaddress]{\texorpdfstring{{}Zhen Liu\corref{mail}}{}}
\ead{liuzhen@impcas.ac.cn}

\author[firstaddress]{\texorpdfstring{{}Valerio Formato\corref{mail}}{}}
\ead{valerio.formato@roma2.infn.it}

\author[firstaddress,thirdaddress]{Muhammad Waqas}

\address[firstaddress]{INFN Sezione di Roma Tor Vergata, 00133 Roma, Italy}
\address[secondaddress]{Institute of Modern Physics (IMP), Chinese Academy of Sciences, Lanzhou, 730000, China}
\address[thirdaddress]{Physics and Astronomy Department, University
of Padua, 35131 Padova, Italy}

\begin{abstract}

The Alpha Magnetic Spectrometer (AMS) aboard the International Space Station provides high-precision measurements of cosmic-ray nuclei fluxes from charge  $Z=1$ to $Z=28$ and beyond. With negligible charge confusion from non-interacting nuclei, the precision of nuclei flux measurements is primarily limited by fragmentation backgrounds originating from heavier cosmic rays interacting within detector materials, particularly between tracker Layers 1 and 2 (L1–L2). As AMS extends its measurements to heavier and rarer nuclei, these fragmentation backgrounds become increasingly dominant, necessitating advanced background suppression methods. To address this challenge, we introduce a Multimodal Domain-Adversarial (MDA) neural network designed to effectively suppress these interaction backgrounds. The MDA model fuses heterogeneous data from the silicon tracker and time-of-flight detectors using specialized sub-networks combined via multi-head attention. Crucially, a domain-adversarial training strategy is employed to learn invariant representations, enabling the model, which is trained  on Monte Carlo simulations, to be reliably applied to flight data. Using phosphorus (P) as a benchmark, we demonstrate its background suppression capabilities. This approach provides a robust, generalizable framework applicable to the measurement of other rare cosmic-ray nuclei with AMS.

\end{abstract}

%\keywords{AMS \and Background supression \and Machine learning \and Domain-Adversarial neural network }

\end{frontmatter}

\section{Introduction}

\subsection{The AMS Detector} \label{subsec:Detector}

\begin{figure}[htbp]
    \centering
    \includegraphics[width=0.65\textwidth]{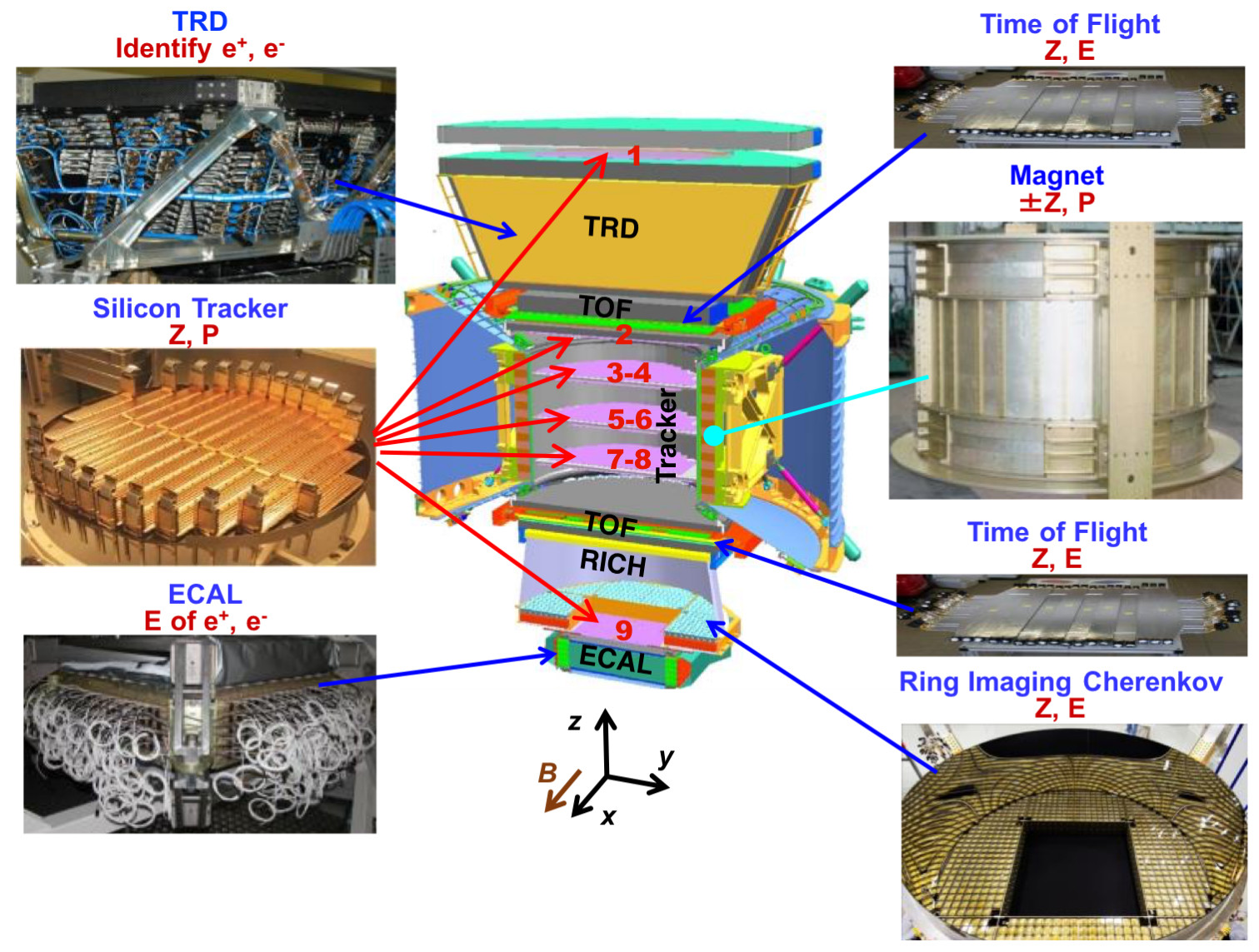}
    \caption{The schematic of the AMS detector.
    \label{fig:detector}}
\end{figure}

The Alpha Magnetic Spectrometer (AMS) is a large-acceptance, long-duration precision magnetic spectrometer operating in space, designed to measure the charge, momentum, and rigidity of cosmic-ray particles~\cite{ams}. The schematic of the AMS detector is shown in   Figure \ref{fig:detector}.  While AMS comprises multiple sub-detectors, the key elements utilized for the background suppression analysis presented in this work are the permanent magnet, the Time of Flight (TOF) scintillation counters, and the precision silicon tracker.

The core of the spectrometer is a \(0.14\)~T permanent magnet~\cite{magnet}. Located above and below the magnet bore are the Upper and Lower TOF counters~\cite{tof}. The TOF system serves multiple crucial functions: it provides a fast charged-particle trigger, determines the flight direction (\(\beta > 0\) for downward-going particles), and measures the particle velocity (\(\beta = v/c\)). Furthermore, the TOF provides a standalone measurement of the absolute charge (\(Z\)) based on the rate of energy loss (\(\mathrm{d}E/\mathrm{d}x \propto Z^2\)).

The precision silicon tracker~\cite{tracker} consists of nine layers of double-sided silicon microstrip detectors. The first layer (L1) is located at the top of the detector, the second (L2) just above the magnet, six layers (L3 to L8) within the magnet bore (constituting the inner tracker), and the final layer (L9) at the bottom. The maximum lever arm from L1 to L9 is approximately 3~m. Operating within the magnetic field, the tracker measures the rigidity (\(R = p/Z\)) of charged particles by precisely reconstructing their trajectories. The coordinate resolution in the bending plane is \(10~\mu\mathrm{m}\) for \(Z = 1\) particles and improves to \(5~\mu\mathrm{m}\) for \(Z = 6\) and heavier nuclei~\cite{tracker_res}. Crucially, each of the nine tracker layers also provides an independent and highly accurate measurement of the absolute charge for incoming particles ranging from \(Z = 1\) up to \(Z = 30\).

\subsection{Heavy nuclei measurement} \label{subsec:nuclei}

Operating on the International Space Station, the AMS has collected 263 billion cosmic-ray events from May 19, 2011, to November 26, 2024. Leveraging this extensive dataset, AMS has previously published precision measurements of cosmic-ray nuclei ranging from protons (\(Z=1\)), helium, and lithium, up to silicon (\(Z=14\)), as well as sulfur (\(Z=16\)) and iron (\(Z=26\)). Recently, these precision measurements have been successfully extended to include phosphorus (\(Z=15\)), chlorine (\(Z=17\)), argon (\(Z=18\)), potassium (\(Z=19\)), and calcium (\(Z=20\))~\cite{ams_pckca}.

A primary challenge in the precise measurement of these heavy nuclei is the effective rejection of fragmentation backgrounds. Cosmic rays can interact with the detector materials, fragmenting into lighter nuclei and mimicking the signal. For instance, phosphorus is significantly less abundant than its heavier neighbor, sulfur. When sulfur nuclei enter the detector, they can undergo nuclear interactions with the materials at the top of AMS or between Tracker L1 and L2. This fragmentation produces secondary phosphorus nuclei, constituting a major background. In general, the dominant sources of such background are the interactions of heavier primary nuclei (e.g., S, Cl, Ar, and Ca) within the AMS materials located above Tracker L2.

In previous AMS analyses~\cite{ams_al, ams_nuclei}, the background resulting from interactions in the materials between L1 and L2 (which include the TRD and the upper TOF) was evaluated using a template fitting method. This approach involves fitting the charge distribution measured by Tracker L1 using charge distribution templates derived from heavier nuclei. While effective for moderately abundant elements, this method faces severe limitations for the upcoming measurements of sub-iron (sub-Fe) nuclei (e.g., \(Z=21\) to \(25\)). Iron (\(Z=26\)) is vastly more abundant than the sub-Fe elements. Consequently, the fragmentation background originating from Fe will be overwhelmingly large compared to the rare sub-Fe signal events. Relying solely on the traditional template fitting method under such extreme low signal-to-background conditions would introduce unacceptably large systematic uncertainties. Pushing the frontier of heavier nuclei measurements strictly requires novel, high-performance signal-background discrimination techniques.

Deep learning methods offer a highly promising avenue for particle physics~\cite{zhen3,zhen2,zhen1}, including this complex discrimination task.
However, a common challenge in applying machine learning to high-energy physics data is the discrepancy between Monte Carlo (MC) simulations and real experimental data.
 Despite a comprehensive and detailed understanding of the detector response and hadronic cross-sections~\cite{YAN2020121712}, residual MC-data inconsistencies inevitably persist, and can compromise the reliability of models trained purely on simulations. To overcome this critical bottleneck, this paper proposes a Multimodal Domain-Adversarial (MDA) approach. By leveraging adversarial learning between MC and flight data, the network is forced to extract domain-invariant features. Using the measurement of phosphorus as a benchmark, we demonstrate that this MDA method achieves high-performance background rejection while ensuring the strict reliability required for precision physics analysis, paving the way for future sub-Fe measurements.

\section{Input features} \label{sec:Input_features}

In the AMS heavy nuclei flux measurements, the key sub-detectors used are silicon tracker layers L1 through L8 (Tracker L1 and Inner Tracker), and the upper time-of-flight scintillation counters (TOF layers 1 and 2). The input features to the MDA network are therefore constructed exclusively from these sub-detectors.

We extract low-level reconstructed quantities, such as per-layer charge, track hit positions, and cluster information, rather than high-level derived variables (e.g., inner tracker combined charge, track fitting quality $\chi^2$), etc. Since high-level variables are computed from the low-level measurements, they carry no additional independent information and are therefore omitted to avoid redundancy.

In addition to these low-level features, we include reconstructed rigidity variables: the L1–inner rigidity (fitted using Tracker layer L1 to L8) and the inner rigidity (fitted using layers L2 to L8). The measured rigidity spans a wide range from 2 GV to 2 TV, over which the detector response exhibits strong kinematic dependencies. Providing these explicit kinematic anchors prevents the network from having to implicitly deduce the particle's momentum, thereby improving its ability to evaluate rigidity-dependent background probabilities.

To ensure that the network does not exploit simulation artifacts, we evaluate the data–MC agreement of each candidate feature. Certain variables, such as the raw energy deposit, are not sufficiently well-tuned in the Monte Carlo simulation and exhibit significant discrepancies with respect to flight data. Such features are excluded from the input set. Although extensive efforts have been devoted to achieving data–MC consistency in the AMS simulation framework~\cite{YAN2020121712}, residual discrepancies in the retained features are unavoidable. This is precisely the motivation for the domain-adversarial training strategy described in the following section, which is designed to learn representations that are invariant to the remaining data–MC differences.

Finally, we examine the multicollinearity among the selected features by computing the Variance Inflation Factor (VIF). Features with VIF $\geq$ 5 are removed to reduce redundancy and improve training stability.

Following the event selection, the final set of input variables is grouped into the following categories:

\paragraph{Tracker charge information}
\begin{itemize}
    \item Layer-by-layer charge measurements from Tracker Layers L2 to L8 for the main track. The charge information from Tracker Layer L1 is intentionally excluded in order to avoid introducing direct dependence on the L1 charge measurement used in the physics analysis.
\end{itemize}

\paragraph{Tracker kinematic information}
\begin{itemize}
    \item L1+Inner rigidity measurements for both the main and secondary tracks.
    \item Inner rigidity measurements for both the main and secondary tracks.
\end{itemize}

\paragraph{Tracker spatial information}
\begin{itemize}
    \item Reconstructed hit positions (X and Y coordinates) for both the main and secondary tracks across Tracker Layers L1 to L8.
\end{itemize}

\paragraph{Tracker cluster topology information}
\begin{itemize}
    \item Layer-by-layer ratios of on-track clusters to the total number of clusters for both X and Y sides across Tracker Layers L1 to L8.
\end{itemize}

\paragraph{Upper TOF information}
\begin{itemize}
    \item TOF $\beta$ measurements, corresponding to the reconstructed particle velocity.
    \item Reconstructed charge measurements from Upper TOF Layers 1 and 2.
    \item Number of on-time TOF clusters in Upper TOF Layers 1 and 2.
    \item Ratios between the number of clusters used for charge reconstruction and the total number of TOF clusters in the upper TOF system.
\end{itemize}

\begin{figure}[htbp]
    \centering
    \includegraphics[width=0.65\textwidth]{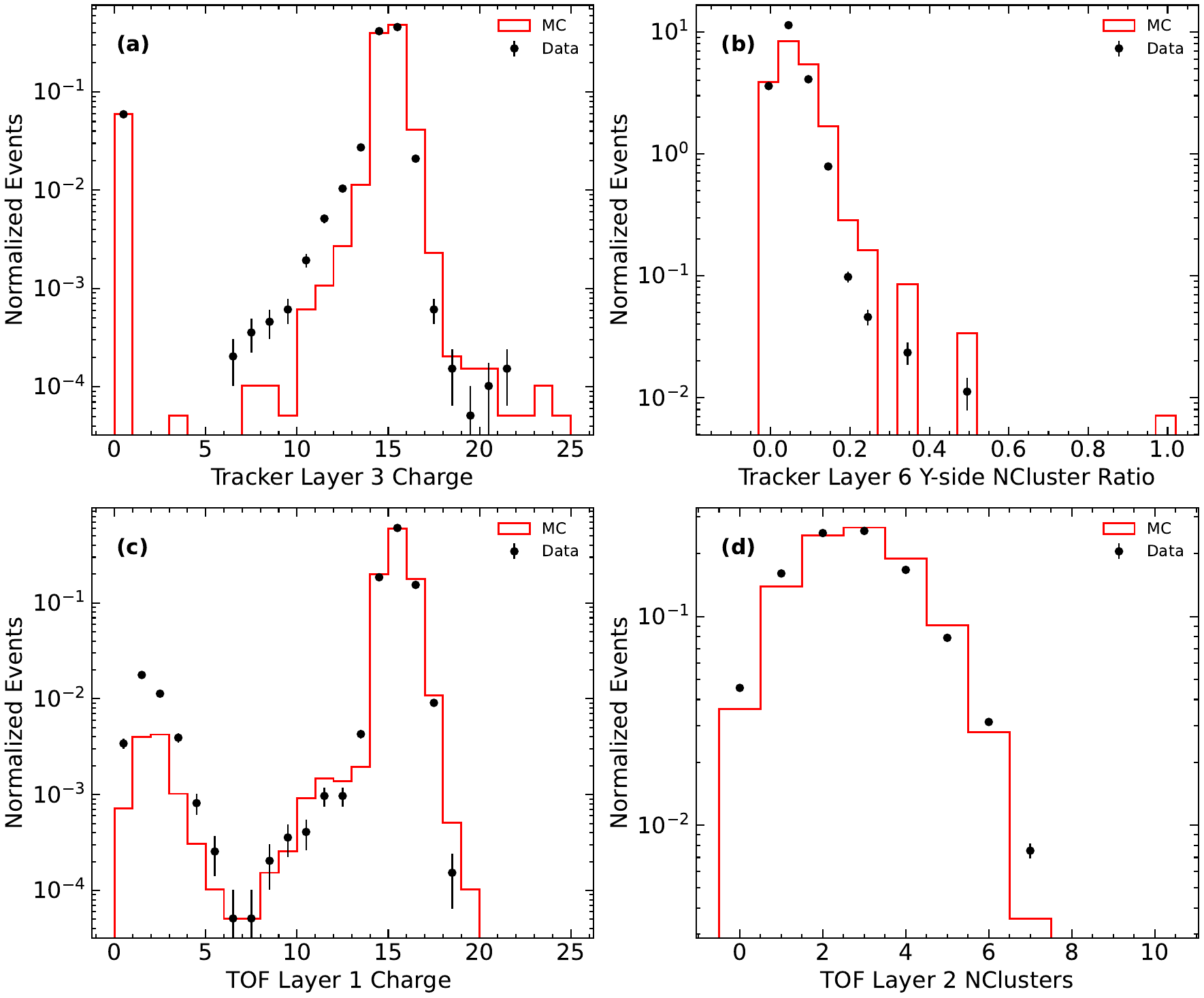}
    \caption{Examples of normalized data and Monte Carlo distributions for four representative input variables in the rigidity interval 20–50 GV.
    \label{fig:compare_20_50GV}}
\end{figure}

Figure \ref{fig:compare_20_50GV} shows examples of data and phosphorus (P)  Monte Carlo (MC) distributions for four representative input variables in the rigidity interval 20–50 GV.  The events are selected with P charge selection, requiring the event to be downward-going and to have a reconstructed track in the inner tracker that passes through L1. As seen, MC generally follows data, while residual discrepancies remain.

\begin{figure*}[htbp]
    \centering
    \includegraphics[width=0.65\textwidth]{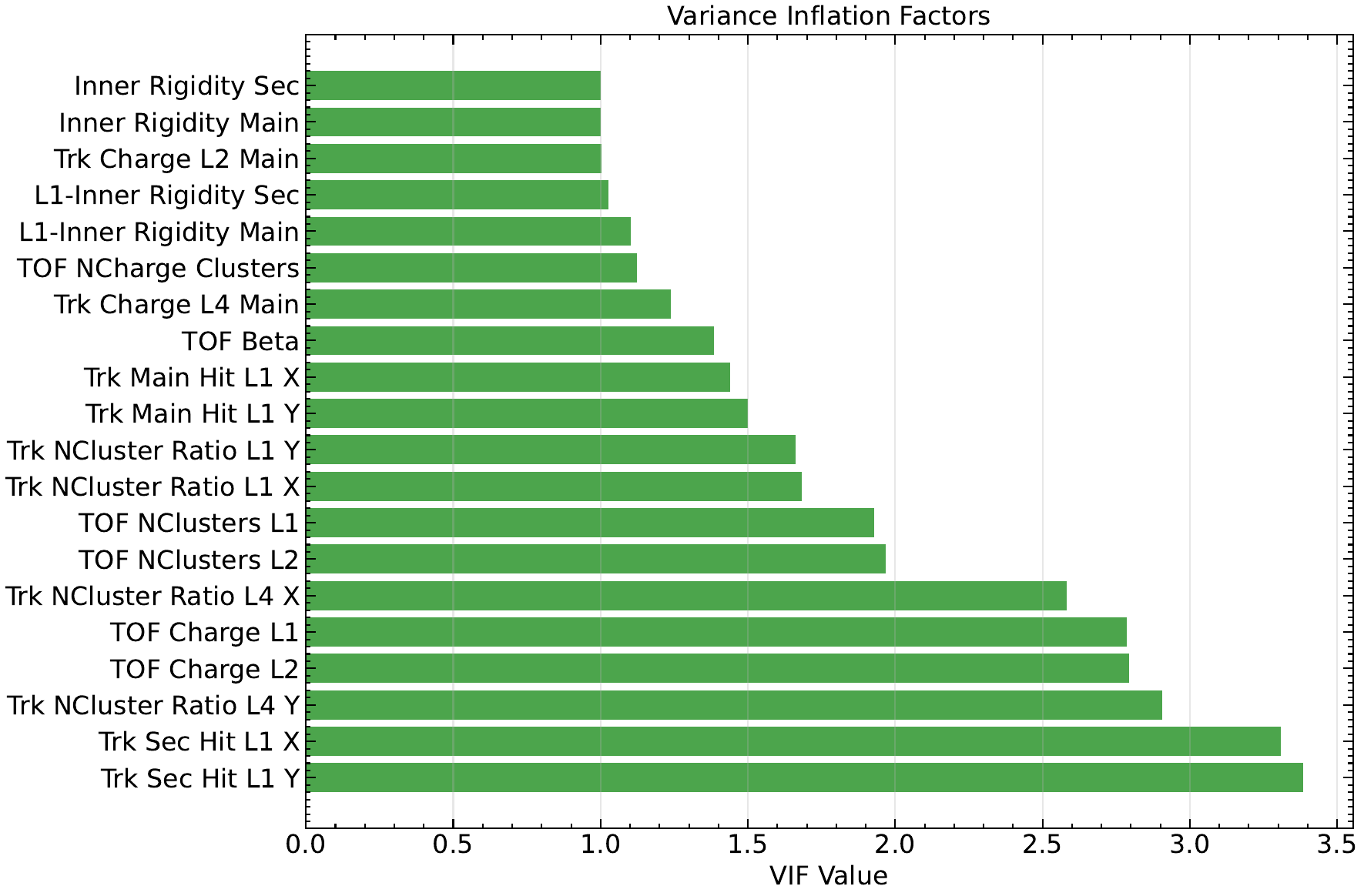}
    \caption{Variance inflation factors (VIFs) for 20 representative variables from the input categories listed in Sec. \ref{sec:Input_features}. Each bar corresponds to an individual variable. The plotted quantities represent several input categories.
    \label{fig:vif}}
\end{figure*}

Figure \ref{fig:vif} shows the VIF for multicollinearity assessment. VIF  was computed for each type of feature. As seen, all VIF values remain below 5, and most are clustered around 1–3.5. This indicates that the retained variables exhibit only mild linear dependence and that no severe redundancy is introduced into the final feature set.

\section{Model} \label{sec:Model}

\begin{figure*}[htbp]
    \centering
    \includegraphics[width=0.85\textwidth]{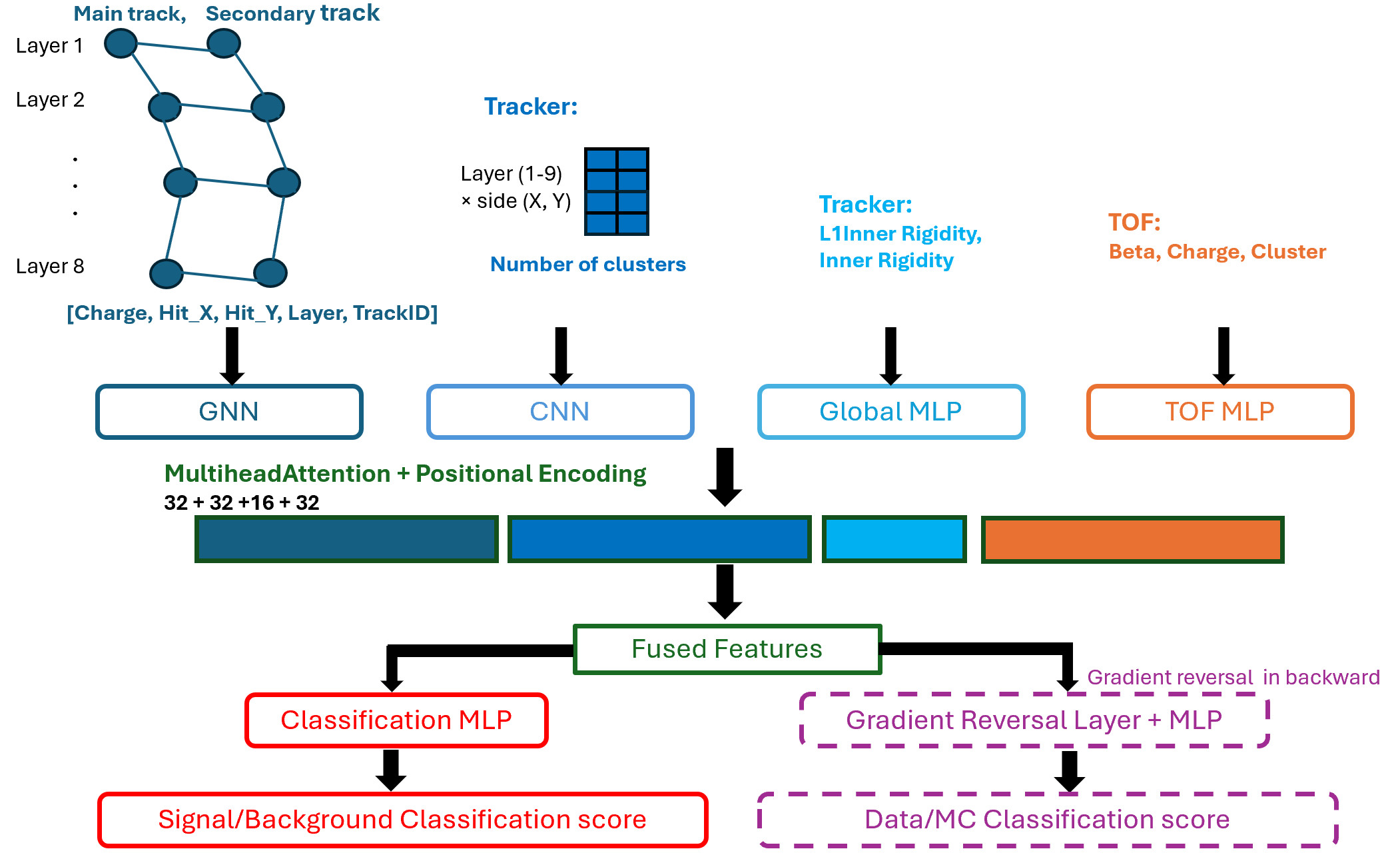}
    \caption{MDA architecture
    \label{fig:MDA_architecture}}
\end{figure*}

This work implements a Multimodal Domain-Adversarial (MDA) neural network designed for signal/background discrimination of heavy nuclei in AMS analyses.
The model combines heterogeneous detector information using several specialized sub-networks. It also reduces discrepancies between simulation and data 
through adversarial training.
The source code is available in the GitLab repository\footnote{\url{https://gitlab.cern.ch/zhenli/mda_for_heavy_nuclei}}, and the overall architecture is illustrated in Figure~\ref{fig:MDA_architecture}.

\subsection{Feature Extraction Modules} \label{sec:feature_extraction}

The feature extractor consists of four specialized sub-networks, each processing a different category of detector information:

\paragraph{Tracker GNN}
A Graph Neural Network (GNN) takes as input the hit positions and charge measurements from both main and secondary tracks across different tracker layers and sides. Each hit is represented as a node with features including charge, hit coordinates, layer index, and track identity. A graph-based representation naturally captures inter-layer correlations and cross-track relationships, making it well suited for encoding the spatial topology of particle interactions in the silicon tracker ~\cite{gnn}.

\paragraph{Tracker CNN}
A Convolutional Neural Network (CNN) processes the cluster information on each tracker layer and side. The cluster ratios are organized in a two-dimensional image-like format with axes corresponding to layer index (1 to 8) and side ($x$, $y$), enabling convolutional layers to learn spatial patterns in the detector response along the particle trajectory ~\cite{cnn}.

\paragraph{Global MLP}
A Multi-Layer Perceptron (MLP) processes event-level scalar variables from the tracker, namely the L1-inner rigidity and the inner rigidity. These global
quantities encode the momentum information of the event and allow the network to learn rigidity-dependent discrimination strategies.

\paragraph{TOF MLP}
A separate MLP handles the Time-of-Flight inputs, including the particle velocity $\beta$, per-layer charge measurements, and cluster multiplicities of
the upper TOF counters. This sub-network captures independent charge and velocity information provided by the TOF system.

Each sub-network produces a fixed-length latent representation (embedding vector) summarizing its corresponding detector response.

\subsection{Multimodal Feature Fusion} \label{sec:fusion}

The embedding vectors from the four sub-networks are fused using a multi-head self-attention mechanism with positional encoding ~\cite{Attention}. Compared to simple concatenation, the attention-based fusion offers several advantages. First, it dynamically reweights each modality on an event-by-event basis, emphasizing the most informative detector signals for each individual event. Second, it enables the model to learn cross-modal correlations automatically, capturing complementary information across sub-detectors that would otherwise require explicit feature engineering ~\cite{multimodal}. The resulting fused representation serves as the input to both the  signal/background classifier and the domain-adversarial branch described below.

\subsection{Signal/Background Classifier} \label{sec:classifier}

The fused representation is passed to a classification MLP that outputs a signal/background score. This classifier is trained on MC
samples in which signal and background events are labeled by their true particle identity. The training objective minimizes the binary cross-entropy
loss between the predicted score and the true label.

\subsection{Domain-Adversarial Training} \label{sec:domain_adversarial}

Because the signal/background classifier is trained entirely on simulation, any discrepancy between MC and flight data may degrade its performance when applied to real events. To address this, a domain-adversarial training strategy is adopted following the principle of  Domain-Adversarial Neural Networks~\cite{JMLR:v17:15-239, McEneaney_2023}.

In addition to the signal/background classifier, a data–MC discriminator is introduced. This discriminator receives the same fused representation and is
trained to distinguish simulated events from real data. A gradient reversal layer is inserted between the fused features and the discriminator: during
back-propagation, the gradients from the discriminator are reversed in sign before they reach the feature extraction modules. As a result, the feature
extractor is trained to simultaneously minimize the signal/background classification loss and maximize the domain discrimination loss. This
adversarial interplay forces the network to learn representations that are invariant to data–MC differences while preserving the physics-relevant features necessary for signal/background separation.

Through this strategy, the model trained on MC simulation can be reliably applied to real data, with the domain-invariant features ensuring robust
generalization across the two domains.

\section{Model Training} \label{sec:Training}

\subsection{Input sample} \label{sec:Input_sample}
The Multimodal Domain-Adversarial (MDA) network is developed for signal and background discrimination in AMS heavy nuclei analyses. This section describes the event selection criteria and the composition of the MC simulation and flight data samples, utilizing the Phosphorus (P) flux measurement as a representative application.

\paragraph{Event Selection}
A preselection procedure is applied to all events to ensure the quality of the reconstructed variables~\cite{ams_pckca}. Basic instrumental and environmental filters are applied, including the removal of bad runs, the South Atlantic Anomaly cut, general data quality cuts, and the geomagnetic cutoff requirement. The standard AMS physical trigger selection is also required.

Events are required to be downward-going and to have a valid reconstructed track in the inner tracker that passes through Tracker L1. 

The charge selection for P candidates is defined using independent measurements from Tracker L1, the Inner Tracker, and the Upper TOF. For L1 measurements, consistency between the X side and Y side charge values is required. An L1 good charge status is also required to select charge signals produced only within the active area of the Silicon Tracker ladders. The analysis is restricted to events with a reconstructed rigidity between \(1\) GV and \(2000\) GV.

\paragraph{Sample Composition}
The model is designed to distinguish true P signal events from heavier nuclei background events such as Sulfur (S) fragmenting into P-like signatures. The datasets are composed of the following samples:

\begin{itemize}
    \item \textbf{MC Signal:} 100,000 simulated P events.
    \item \textbf{MC Background:} 100,000 simulated S background events passing the aforementioned P selections. Although the network is trained exclusively with S as the background, it demonstrates robust generalization capabilities when subsequently applied to a broader range of nuclear species.
    \item \textbf{Flight Data:} 160,000 flight data events passing the selection criteria.
\end{itemize}

The MC samples are randomly split, with 80 percent used for training and 20 percent reserved for testing. To prevent the network from learning underlying kinematic differences between the datasets, the MC events are filtered to match the rigidity dependence of the flight data. Specifically, the L1+Inner Rigidity distribution of the MC samples is aligned to be consistent with the data. The nominal MC training sample contains equal numbers of phosphorus signal and sulfur-induced background events, while the selected flight sample is enriched in phosphorus. At fixed rigidity, different signal and background fractions can provide a cue to the domain discriminator. Adversarial training may therefore suppress some information useful for signal--background separation. The flight-data results in Sec.~6.1 correspond to the training composition used in this study. Their sensitivity to this composition remains to be evaluated.

\subsection{Training} \label{sec:training}

The model is implemented in PyTorch and trained on a single NVIDIA Tesla T4
GPU. The training dataset contains  320k events, yielding an average training time of about 1 minute per epoch. The AdamW optimizer is used with a one-cycle learning rate schedule, reaching a maximum learning rate of $10^{-3}$. Both the signal/background classification loss and the data–MC domain discrimination loss are formulated as binary cross-entropy with logits. The gradient reversal layer (GRL) weight $\lambda$ is linearly ramped up to $\lambda_{\max} = 1.5$ over the course of training, following the scheduling strategy proposed in ~\cite{JMLR:v17:15-239}, to allow the feature extractor to first learn discriminative representations before the adversarial constraint is fully activated. Spectral normalization is applied to the domain classifier to stabilize the adversarial training dynamics.

An early stopping procedure is employed by monitoring the validation loss, which begins to converge around epoch 30 while the training loss continues to
decrease slowly. The model snapshot at epoch 32, corresponding to the minimum validation loss, is selected as the final model. The full set of hyperparameters is summarized in Table~\ref{tab:Hyperparameters}.
\begin{table*}[htbp]
\centering
\caption{Model hyperparameters. Some parameters were tuned via a basic grid search.}
\label{tab:Hyperparameters}
\renewcommand{\arraystretch}{1.3}
\begin{tabular}{ll}
\toprule
\textbf{Parameter} & \textbf{Value} \\
\midrule
\multicolumn{2}{c}{\textit{Domain Adversarial}} \\
\midrule
Activation function & LeakyReLU (\(\text{negative slope} = 0.2\)) \\
Dropout rate & 0.1 \\
Domain classifier depth & Medium (4 fully connected layers) \\
Max GRL weight (\(\lambda_{\max}\)) & 1.5 \\
Spectral Normalization & Applied to Domain Classifier \\
\midrule
\multicolumn{2}{c}{\textit{Training \& Optimization}} \\
\midrule
Optimizer & AdamW (\(\beta_1 = 0.9, \beta_2 = 0.999\)) \\
Weight decay & \(10^{-3}\) \\
Learning rate strategy & One-cycle policy \\
Max learning rate & \(10^{-3}\) \\
Initial / Final division factor & 25 / 50 \\
Task / Domain Loss & BCEWithLogitsLoss \\
Training epochs & 40 \\
Batch size & 2048 \\
Gradient clipping & 0.3 \\
\bottomrule
\end{tabular}
\end{table*}

\subsection{Training results}

\begin{figure*}[htbp]
    \centering
    \includegraphics[width=0.85\textwidth]{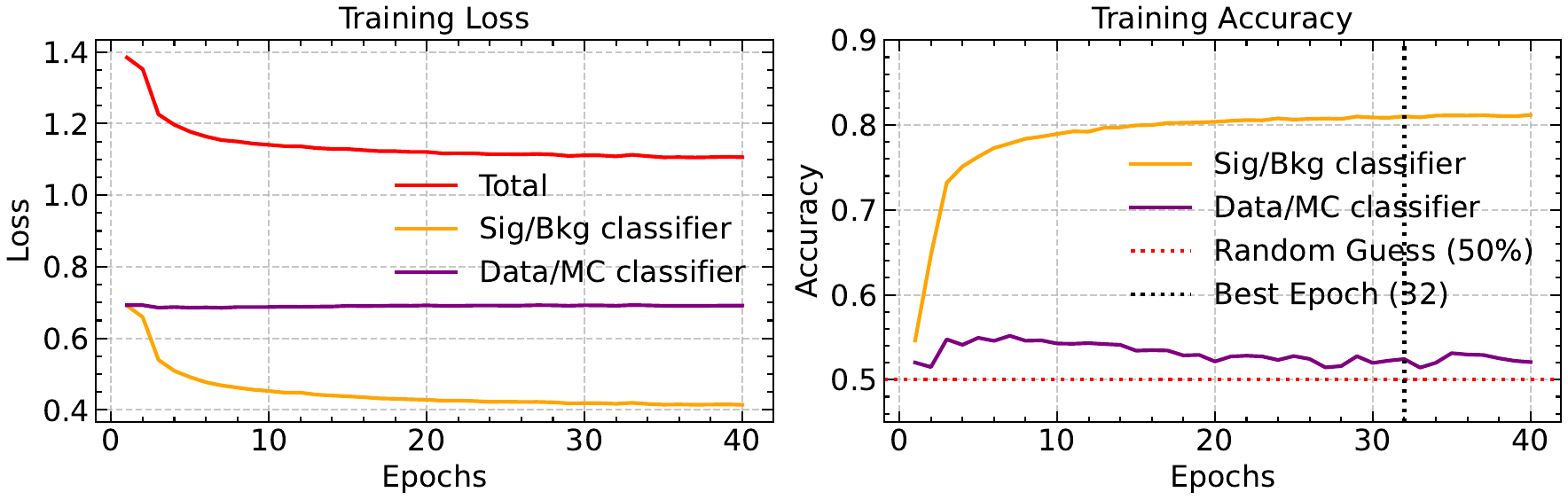}
    \caption{Training loss (left) and accuracy (right) as a function of training epoch. The total loss (red), signal/background classifier loss (orange), and data–MC classifier loss (purple) are shown in the left panel. The corresponding accuracies for the signal/background classifier (orange) and the data–MC classifier     (purple) are shown in the right panel, along with the 50\% random-guess baseline (red dotted). The vertical dotted line indicates the best epoch selected based on validation performance.}
    \label{fig:training_overview}
\end{figure*}

\begin{figure*}[htbp]
    \centering
    \includegraphics[width=0.85\textwidth]{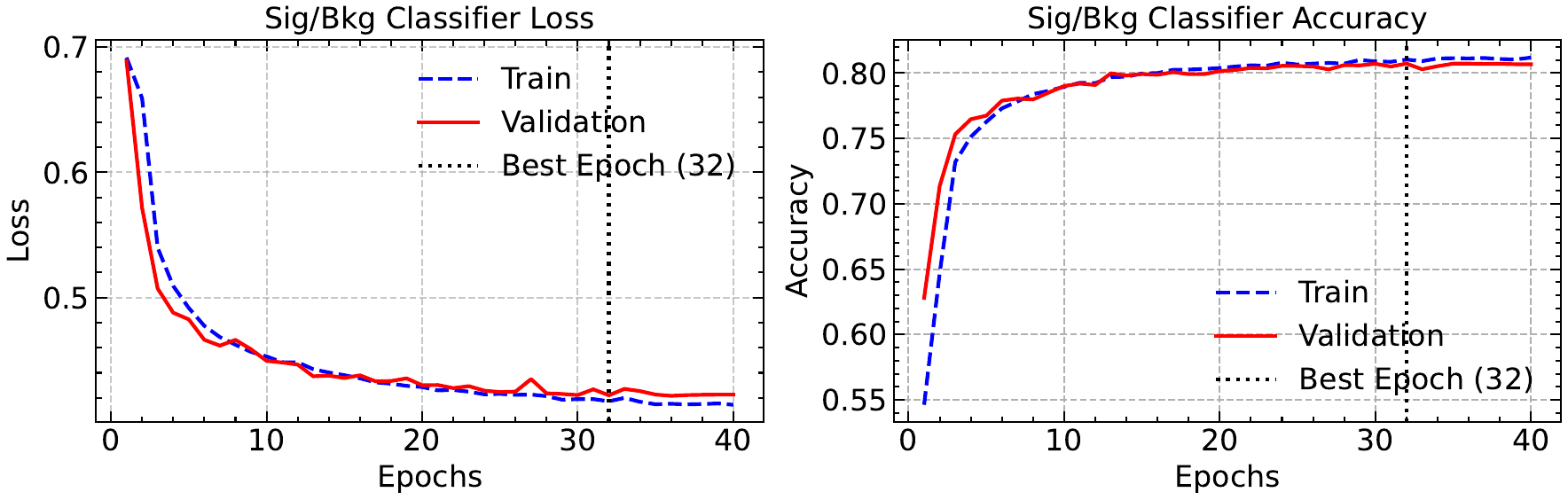}
    \caption{Signal/background classifier loss (left) and accuracy (right) for the training and validation datasets as a function of training epoch. Both the loss and accuracy curves exhibit consistent behavior between the training and validation samples throughout the training process.}
    \label{fig:sigbkg_trainval}
\end{figure*}

The training results are summarized in Figures~\ref{fig:training_overview} and ~\ref{fig:sigbkg_trainval}.

Figure~\ref{fig:training_overview} shows the training loss and accuracy as a function of epoch for the signal/background classifier, the data–MC domain classifier, and the total combined loss. The signal/background classifier loss decreases steadily and converges after approximately 30 epochs, while its accuracy reaches $\sim$80\%. The data–MC classifier loss remains stable around $\sim$0.69 throughout the training, consistent with the binary cross-entropy of a random classifier, $-\ln(0.5) \approx 0.693$.
Correspondingly, the data–MC classifier accuracy fluctuates between 51\% and 54\%, only marginally above the 50\% random-guess baseline.
The area under the receiver operating characteristic curve (AUC) for the domain discriminator was 0.515 on the events used for adversarial training. This value is close to the random-guess expectation of 0.5.
We report this value as a training-sample diagnostic.

Figure~\ref{fig:sigbkg_trainval} shows the signal/background classifier loss and accuracy evaluated separately on the training and validation sets as a function of epoch. Both the training and validation curves exhibit consistent behavior throughout the training process, with no significant divergence observed between them. The validation loss continues to decrease in parallel with the training loss, and the validation accuracy closely tracks the training accuracy, reaching $\sim$80\% at convergence. The absence of a
notable gap between the training and validation metrics indicates that the model does not suffer from overfitting, and the learned classification performance generalizes well to unseen data.

\begin{figure*}[htbp]
    \centering
    \includegraphics[width=0.65\textwidth]{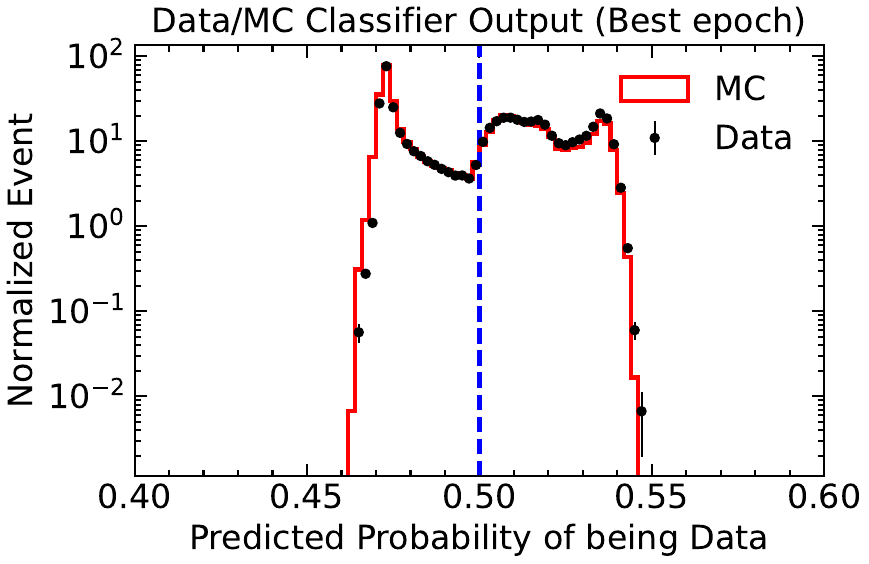}
    \caption{Output score distribution of the data-MC domain classifier at the best epoch. The predicted probability of being classified as data is shown for MC (red curve) and data (black points) samples. The blue dashed line indicates the random-guess threshold at $0.5$. The close agreement between the two distributions demonstrates the effectiveness of the adversarial domain adaptation.}
    \label{fig:domain_score}
\end{figure*}

Additionally, Fig.~\ref{fig:domain_score} shows the output distribution of the domain discriminator for data and MC at the best training epoch. The predicted probability of an event being classified as data is shown for both the data (black points) and MC (red curve) samples. The two distributions overlap closely and are confined to a narrow range around the random-guess value of 0.5.  The events shown in Fig.~\ref{fig:domain_score} were drawn from the adversarial training loader. We also compared the final signal-classifier score distributions for a tight phosphorus control sample in flight data and a truth-tagged phosphorus MC validation sample.
 
\section{Performance on Global MC} \label{sec:MC_Performance}

\begin{figure}[htbp]
    \centering
    \includegraphics[width=0.65\textwidth]{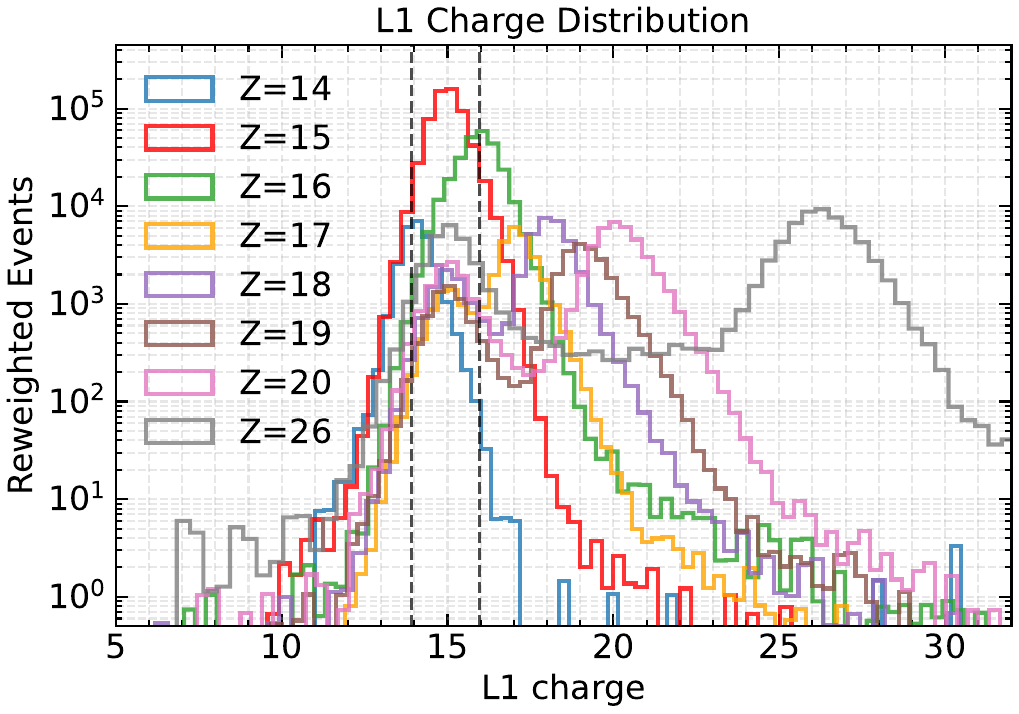}
    \caption{The L1 charge distribution of the reweighted global MC sample. The dashed vertical lines indicate the standard L1 charge cut range applied to select Phosphorus candidates.}
    \label{fig:MC_L1_charge}
\end{figure}

To evaluate the model under realistic conditions with MC sample, a “global MC'' sample was constructed using simulated events for Si (Z=14), P (Z=15), S (Z=16), Cl (Z=17), Ar (Z=18), K (Z=19), Ca (Z=20), and Fe (Z=26). These samples were subjected to the standard P event selection criteria, with the exception of the Tracker L1 charge cut. To reproduce their natural relative abundances, the events were reweighted according to their measured fluxes. Figure~\ref{fig:MC_L1_charge} shows the resulting L1 charge distribution of this global MC sample, where the dashed vertical lines indicate the L1 charge cut range applied to select P candidates.

\begin{figure}[htbp]
    \centering
    \includegraphics[width=0.65\textwidth]{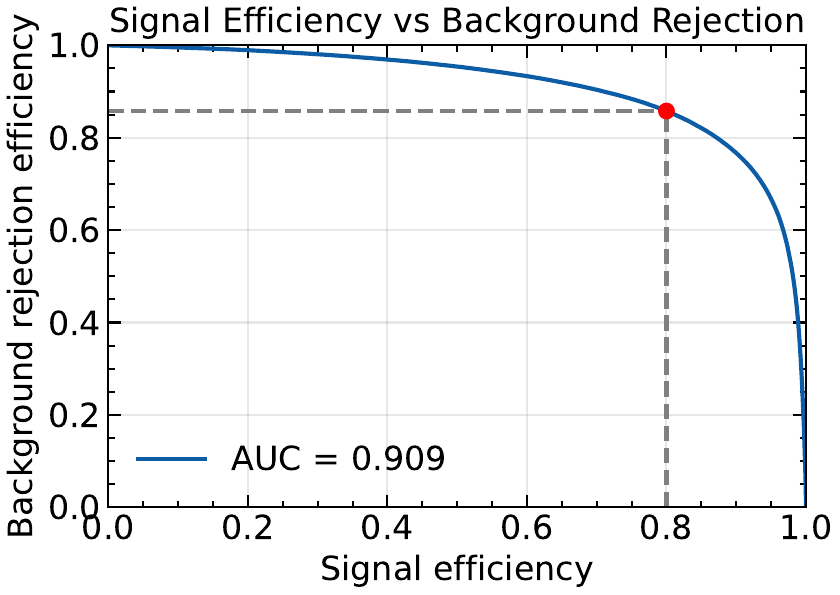}
    \caption{ROC curve of the classifier showing signal efficiency versus background rejection efficiency. The red marker indicates a rigidity-independent score threshold that retains approximately 80\% of the phosphorus signal and rejects approximately 86\% of the background. The background rejection is one minus the fraction of background events passing the MDA cut.}
    \label{fig:MC_roc_curve}
\end{figure}

Figure~\ref{fig:MC_roc_curve} presents the Receiver Operating Characteristic (ROC) curve of the classifier, illustrating the trade-off between signal efficiency and background rejection. For demonstration and visualization purposes, a simple rigidity-independent working point is selected, corresponding to a signal efficiency of 0.8. At this threshold, the model achieves an overall background rejection efficiency of approximately 0.86. This working point is not intended to represent the optimal operating point for a physics analysis. In practice, the MDA score threshold can be further optimized, potentially as a function of rigidity, to achieve the required balance between signal efficiency and background suppression.

\begin{figure*}[htbp]
    \centering
    \includegraphics[width=0.47\textwidth]{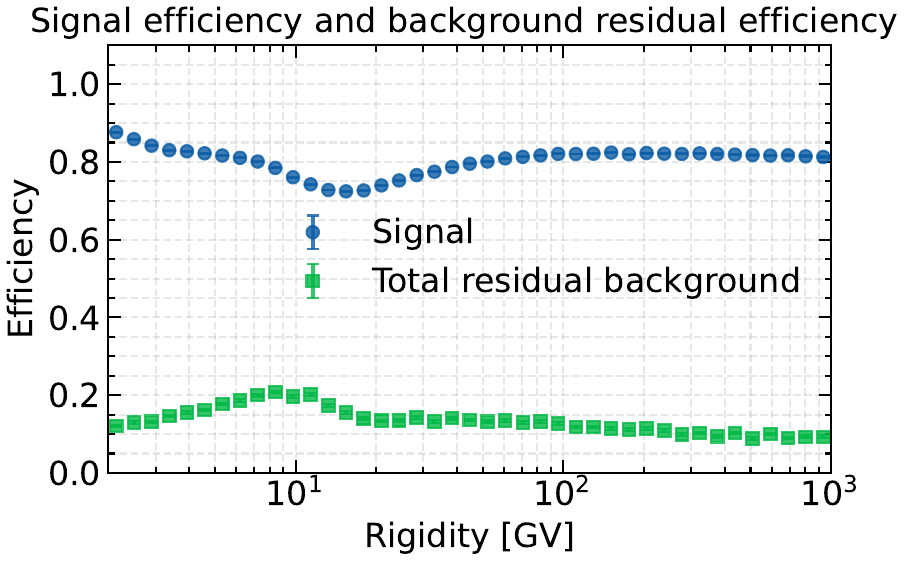}
    \hfill
    \includegraphics[width=0.51\textwidth]{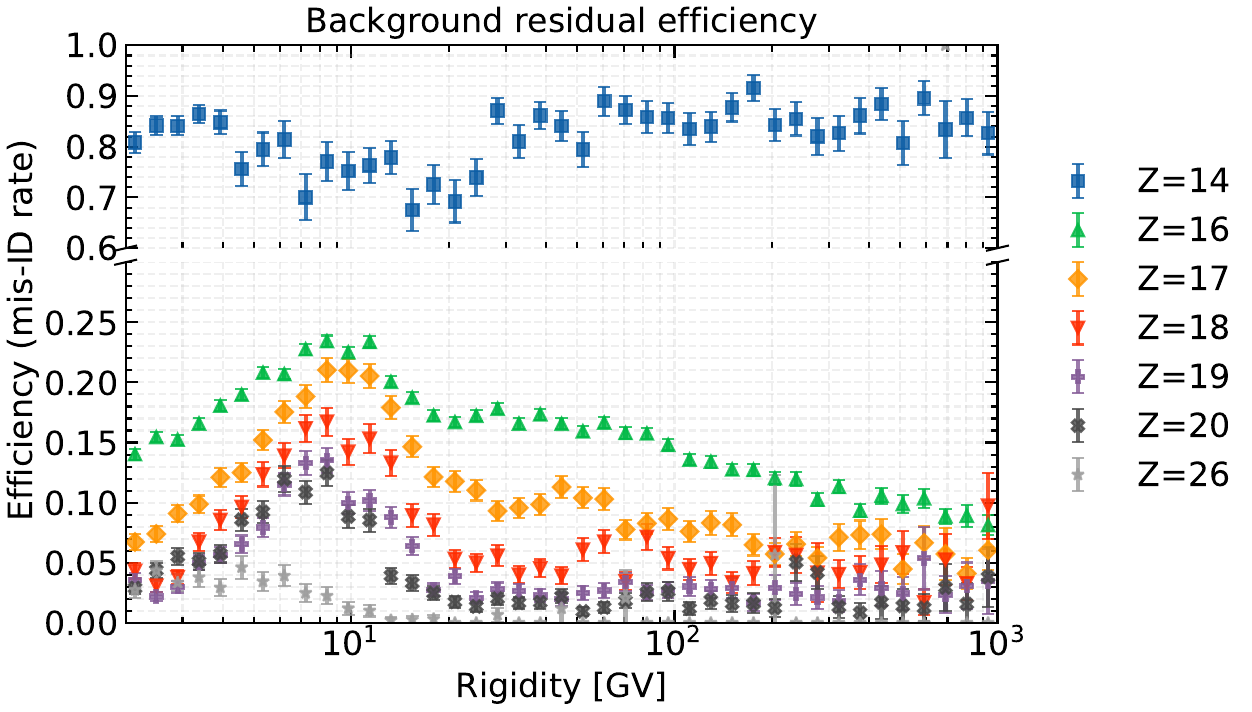}
    \caption{Left: Rigidity dependence of the overall signal efficiency and total residual background efficiency obtained with the chosen rigidity-independent threshold. Right: Breakdown of the residual background efficiency by element. Although the model's output score threshold is continuously tunable, for demonstration purposes, we apply a fixed, rigidity-independent MDA score threshold here that corresponds to an overall signal efficiency of 80\%.}
    \label{fig:MC_combined_eff}
\end{figure*}

Applying this baseline threshold, Figure~\ref{fig:MC_combined_eff} (left) displays the rigidity dependence of the overall signal efficiency and the total residual background efficiency. While a constant, simple threshold yields some rigidity dependence in the efficiencies, in a dedicated physics analysis, this threshold can be finely tuned as a function of rigidity to achieve a flat efficiency response.

Figure~\ref{fig:MC_combined_eff} (right) provides a detailed breakdown of the background residual efficiency for the various background species. The primary observation is the network's strong capability to identify and reject background events from elements heavier than P. These backgrounds primarily arise from fragmentation occurring above Tracker L2, and the model effectively recognizes and suppresses these interacting events. As a secondary effect, the model exhibits a relatively lower discrimination power against Si (\(Z=14\)). This behavior is physically expected, as Si backgrounds largely consist of non-interacting particles that do not undergo fragmentation, thereby leaving signatures very similar to the primary P signal.

Most notably, in the phosphorus MC benchmark, background rejection is generally stronger for heavier parent nuclei (Fig.~\ref{fig:MC_combined_eff}). At the illustrative threshold in Fig.~\ref{fig:MC_combined_eff}, rejection of the simulated iron background exceeds 95\% in many of the displayed rigidity bins. The labeled MC training sample included only phosphorus signal and sulfur-induced background. In sub-iron flux measurements, Fe$\to$Mn and Fe$\to$Cr involve small charge losses close in magnitude to that of S$\to$P. The larger charge losses in Fe$\to$V, Fe$\to$Ti, and Fe$\to$Sc ($\Delta Z = 3, 4, \text{ and }5$, respectively) may give rise to more distinctive fragmentation signatures, offering the potential for substantially improved background rejection in these channels.

\section{Performance on flight data}

\subsection{Background between Tracker L1 and L2}
In the AMS heavy-nuclei flux measurement, the dominant background arises from charge-changing interactions occurring between tracker layers~1 and~2. This background is estimated using a template fit method that exploits the symmetric charge response of the two layers.

Tracker layers~1 and~2 are constructed from the same type of silicon strip sensors and operate under identical conditions, resulting in similar charge measurement responses. However, the two layers play different roles in background estimation. Layer~1 is the topmost tracking layer in AMS, with no additional tracker measurement above it; consequently, a nucleus that undergoes a charge-changing interaction between layers~1 and~2 will register its original charge on layer~1 and a different charge on layers~2 through~8. In contrast, the charge measured on layer~2 can be obtained in a clean manner by applying tight charge selections simultaneously on layer~1 and layers~3 through~8, effectively constraining the particle identity from both sides and suppressing contamination from interactions above or below layer~2.

\begin{figure*}[htbp]
    \centering
    \includegraphics[width=0.42\textwidth]{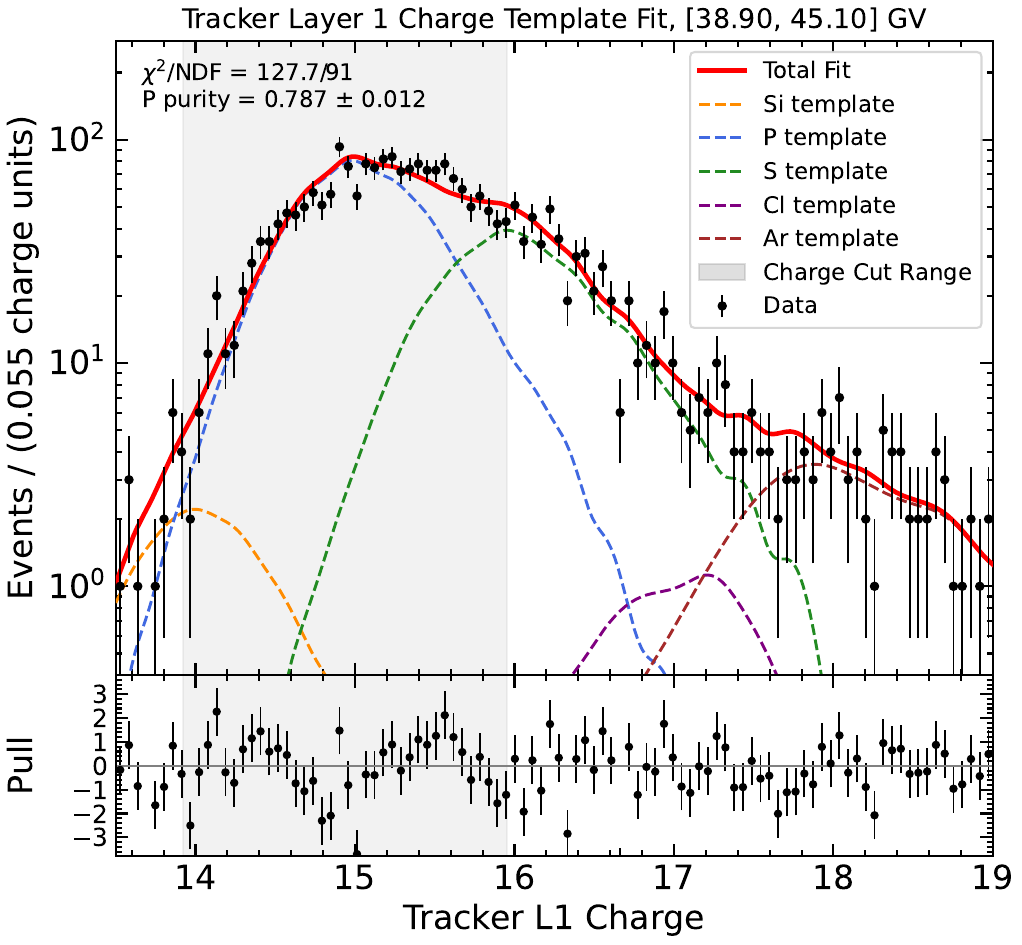}
    \includegraphics[width=0.42\textwidth]{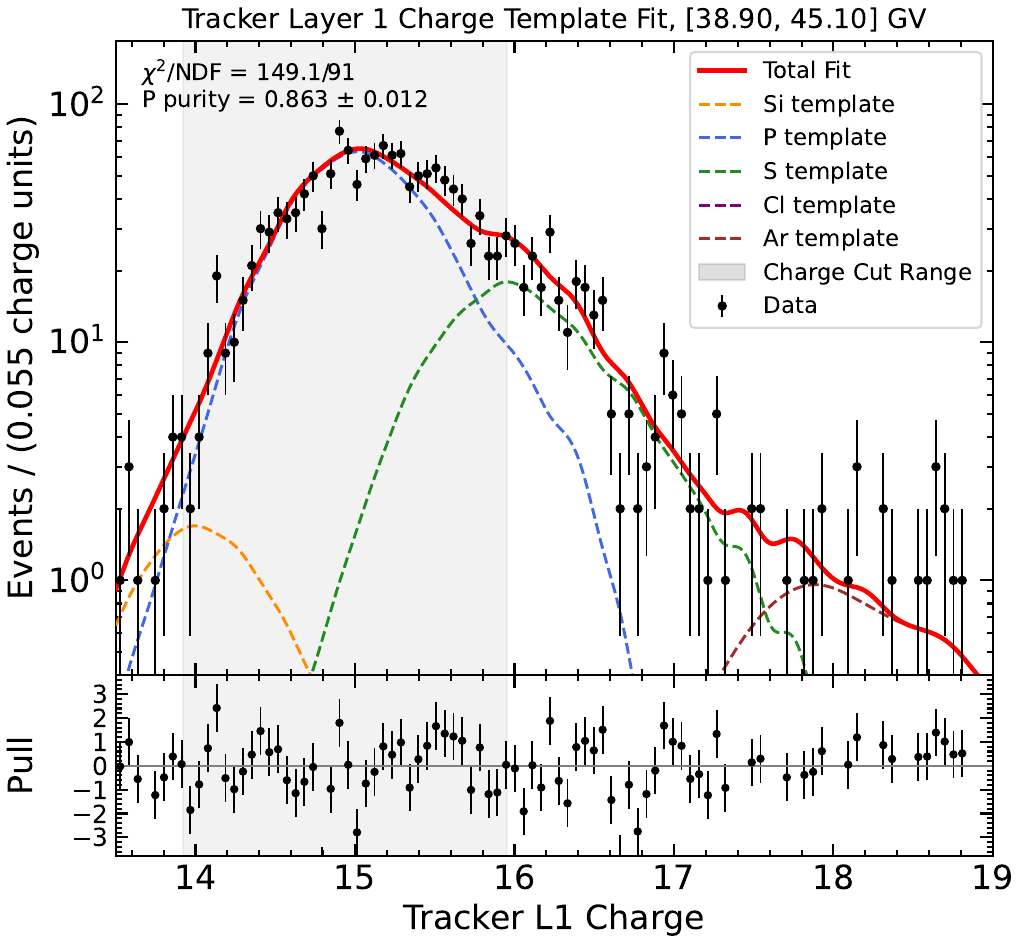}
    \caption{Layer~1 charge distributions of phosphorus candidates in a
representative rigidity bin, fitted with charge templates for silicon (orange),
phosphorus (blue),  sulfur (green), chlorine (magenta), and argon (brown) constructed from layer~2 measurements. The black dots represent the observed layer~1 charge distribution, and the red curve shows the total fit. The shaded band indicates the layer~1 charge selection window applied to select phosphorus events. \textit{Left:} distribution before the MDA selection, showing substantial contamination from silicon and sulfur within the selection window. \textit{Right:} distribution after applying an MDA working point corresponding to a signal efficiency of $\sim$80\%. This selection reduces the background while effectively preserving the primary phosphorus signal.}
     \label{fig:template_fit_example}
\end{figure*}

For the phosphorus ($Z=15$) measurement, the charge distribution observed on layer~1 contains contributions not only from true phosphorus events but also
from neighboring nuclei, primarily silicon ($Z=14$) and sulfur ($Z=16$), whose fragments mimic a phosphorus charge on layer~1 after interacting between the
two layers. To estimate this background, pure charge templates for silicon, phosphorus, sulfur,  chlorine, and argon are constructed from the layer~2 charge distributions obtained with the tight selection described above. The layer~1 charge distribution is then fitted as a linear combination of these templates,
yielding the signal and background fractions. The left panel of Figure~\ref{fig:template_fit_example} shows an example of the template fit result in a
representative rigidity bin. As seen, significant contamination from silicon and sulfur remains within the phosphorus charge selection window.

The MDA network provides a complementary, event-by-event approach to background suppression. The MDA output score is applied as a final selection criterion in the phosphorus event selection chain.

In contrast to the illustrative ROC point used on global MC, the flight data analysis adopts a working point defined directly on the MDA output score.
For demonstration, we use a simple rigidity-independent working point that retains about 80\% of the signal. This choice is close to the operating point that would be adopted in a realistic physics analysis.
The right panel of Figure~\ref{fig:template_fit_example} shows the corresponding layer~1 charge distribution after applying this cut. The background from heavier nuclei is substantially reduced while the phosphorus signal is largely preserved, demonstrating the effectiveness of the domain-adversarial approach in improving the signal purity. Both fits use the same set of L2-derived charge templates, following the L2-template approach of Ref.~\cite{ams_pckca}.

\begin{figure}[htbp]
    \centering
    \includegraphics[width=0.65\textwidth]{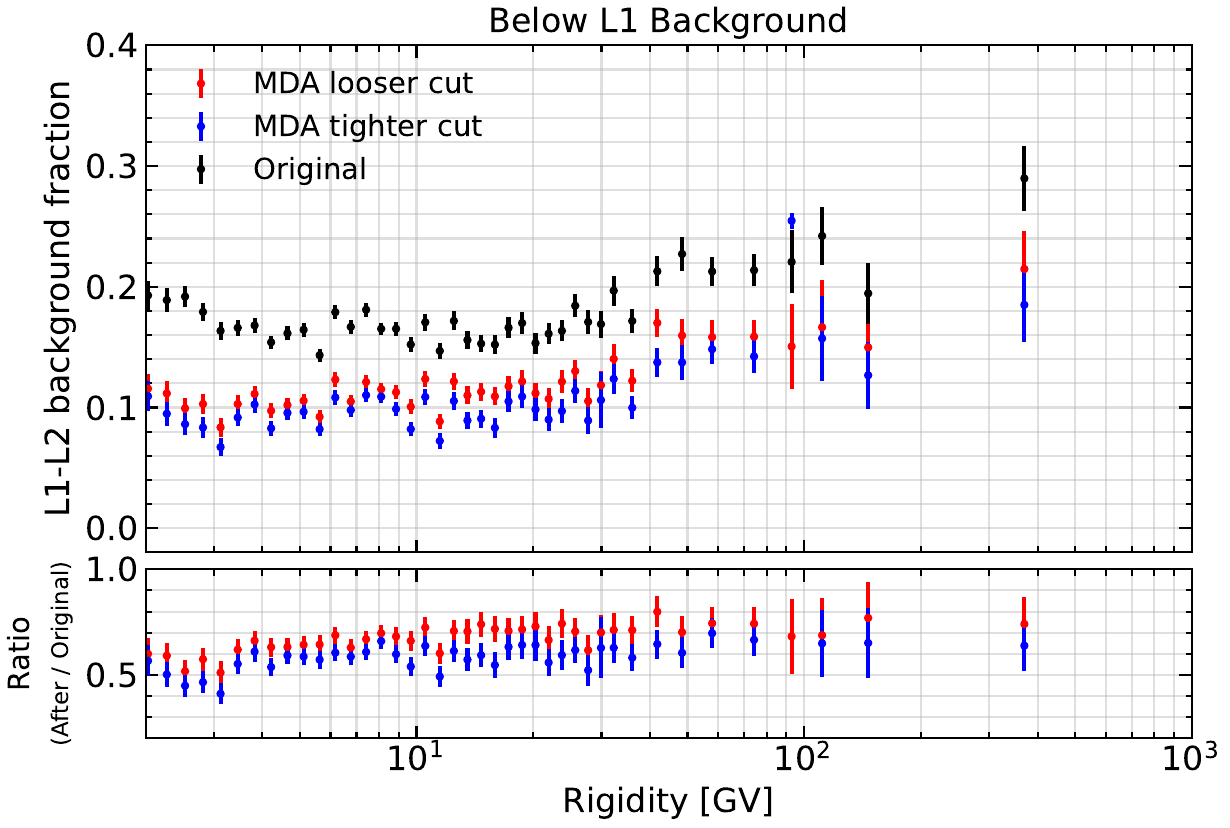}
    \caption{The upper panel shows the L1--L2 background fraction among selected phosphorus candidates before and after the MDA cuts. The background fraction is defined as one minus the phosphorus purity obtained from the L1 charge-template fit. In each rigidity bin, the lower panel shows the background fraction after each MDA cut divided by the fraction in the original sample without an MDA cut. The looser and tighter cuts use rigidity-independent score thresholds and correspond to signal efficiencies of approximately 90\% and 80\%, respectively.}
    \label{fig:template_fit_background}
\end{figure}

Figure~\ref{fig:template_fit_background} shows the background between Tracker L1 and L2,  before and after applying the ML selection. Two rigidity-independent MDA score thresholds are chosen for demonstration: a tighter cut corresponding to a signal efficiency of $\sim$80\%, and a looser cut corresponding to $\sim$90\%. It should be noted again that in a practical physics analysis, a rigidity-dependent threshold is typically applied to maintain a flat and stable selection efficiency across the entire rigidity.

\subsection{Efficiency}

The selection efficiency of the primary P signal, after applying the Machine Learning (ML) cut, is calculated as follows:

\[
\epsilon = \frac{N^{\mathrm{MLcut}}_{\mathrm{selected}}(1-f^{\mathrm{MLcut}}_{\mathrm{AboveL1}})(1-f^{\mathrm{MLcut}}_{\mathrm{BelowL1}})}{N_{\mathrm{selected}}(1-f_{\mathrm{AboveL1}})(1-f_{\mathrm{BelowL1}})} 
\]

\noindent where \(N_{\mathrm{selected}}\) is the number of events passing the baseline P selections. The term \(f_{\mathrm{BelowL1}}\) represents the fraction of fragmentation background originating between Tracker L1 and L2. As discussed comprehensively in previous sections, this is the primary background component that the current MDA network is designed to identify and suppress. 

The term \(f_{\mathrm{AboveL1}}\) denotes the fragmentation background originating from materials above Tracker L1. While the proposed method also effectively rejects a portion of this background, quantifying this rejection power directly from data is generally not possible. Both \(f_{\mathrm{AboveL1}}\) and its post-cut counterpart, \(f^{\mathrm{MLcut}}_{\mathrm{AboveL1}}\), must rely on MC simulations. Although the MDA method significantly reduces data-MC discrepancies, applying an identical score threshold still yields slightly different classification performances between data and MC. This residual difference arises from the limitations of domain adaptation in perfectly aligning high-dimensional distributions. Consequently, \(f^{\mathrm{MLcut}}_{\mathrm{AboveL1}}\) cannot be reliably determined from MC alone.

To ensure a robust analysis, we adopt a deliberately conservative treatment of this background component. For the nominal efficiency estimate, we conservatively assume that the above-L1 background fraction is unchanged by the MDA cut. As a sensitivity check, we set the post-cut fraction to half its pre-cut value to assess how a substantial reduction in this background would affect the inferred phosphorus selection efficiency.

\begin{figure}[htbp]
    \centering
    \includegraphics[width=0.65\textwidth]{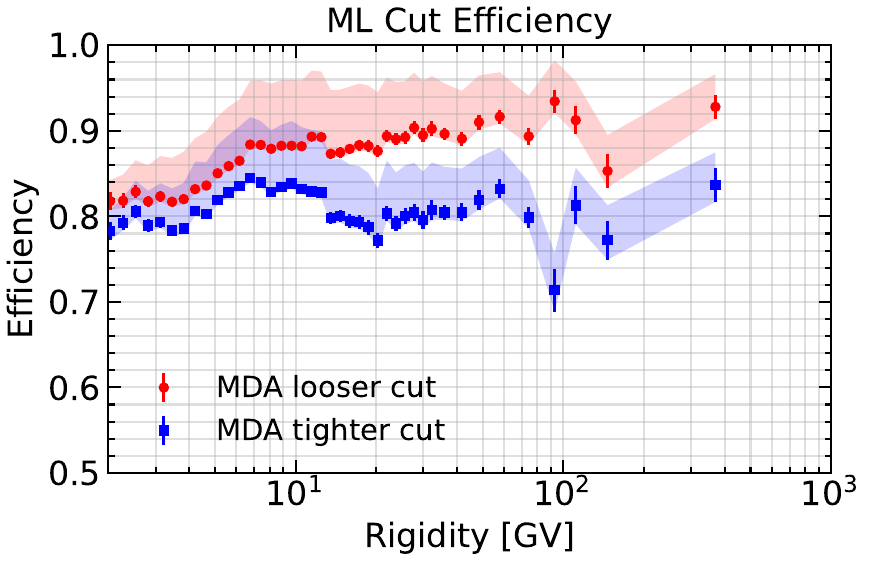}
    \caption{Selection efficiency of the primary phosphorus signal as a function of rigidity. Two rigidity-independent MDA score thresholds are chosen for demonstration purposes: a tighter cut corresponding to a signal efficiency of \(\sim\)80\%, and a looser cut corresponding to \(\sim\)90\%. The error bars include the statistical error and uncertainty of the template-fit background. The shaded bands also include the one-sided variation due to the assumed post-cut above-L1 background fraction.
    As previously noted, in a practical physics analysis, a rigidity-dependent threshold is typically applied to maintain a flat and stable selection efficiency across the entire rigidity spectrum.}
    \label{fig:data_efficiency}
\end{figure}

Under this assumption, a sufficiently large and conservative one-sided systematic uncertainty is assigned to account for the assumed variation in the above-L1 background fraction. The resulting selection efficiency, derived under these conditions, is presented in Figure~\ref{fig:data_efficiency}. The vertical error bars indicate the statistical error and uncertainty of the template-fit background, while the shaded bands represent the total uncertainty including the one-sided systematic contribution described above.

\section{Summary and Conclusions}

In this paper, we present a Multimodal Domain-Adversarial (MDA) neural network specifically developed for the Alpha Magnetic Spectrometer (AMS) to effectively suppress fragmentation backgrounds. The MDA architecture fuses heterogeneous data from the silicon tracker and Time-of-Flight detectors using specialized sub-networks integrated via a multi-head attention mechanism. Crucially, by employing a domain-adversarial training strategy, the network learns domain-invariant representations. This allows the model, which is trained exclusively on MC simulations, to be reliably and directly applied to real flight data, successfully mitigating the inherent data-MC discrepancies.

Using phosphorus (P) as a benchmark, we demonstrate the strong background suppression capabilities and the excellent generalization potential of the model. Although the network was trained using only the adjacent heavier element sulfur (S) as the background source, it exhibits an even stronger rejection power against fragmentation backgrounds originating from heavier nuclei such as Iron (Fe). Application to real flight data reveals that when implemented as a final selection cut, the MDA model achieves a substantial background reduction by roughly  35\% to 50\% depending on rigidity, while maintaining a signal efficiency of 80\% to 90\%.

These results highlight the robustness and scalability of the proposed MDA approach. Furthermore, this methodology provides a promising tool for future AMS measurements of sub-iron (sub-Fe) nuclei, where precise background rejection from the highly abundant iron (Fe) group is of critical importance.

The rigidity-independent score thresholds used here illustrate the performance of the method in the phosphorus benchmark. A dedicated sub-iron analysis should assess the sensitivity to rigidity-dependent changes in the signal fraction and the mixture of fragmentation backgrounds, and validate the choice of operating point as a function of rigidity.

\section*{Acknowledgements}
    Research supported by:     \href{http://dx.doi.org/10.13039/501100004007}{INFN}  and \href{http://dx.doi.org/10.13039/501100003981}{ASI} under ASI-INFN agreement No.\ 2021-43-HH.0,

\end{document}